\documentclass[aps,prl,twocolumn,groupedaddress,draft,showpacs]{revtex4}
\usepackage{graphicx,epsf}
\newif\ifdraft
\drafttrue
\newif\ifpreprint
\preprinttrue

\def\fig#1{fig.~{\ref{#1}}}
\def\Fig#1{Fig.~{\ref{#1}}}

\def\tab#1{table~{\ref{#1}}}

\def\spa#1.#2{\left\langle#1\,#2\right\rangle}
\def\spb#1.#2{\left[#1\,#2\right]}
\def\tree{{\rm tree}}

\def\nn{\nonumber}

\def\eqn#1{eq.~(\ref{#1})}

\def\NeqFour{{{\cal N}=4}}

\def\NeqEight{{{\cal N}=8}}

\def\be{\begin{equation}}
\def\ee{\end{equation}}
\def\bea{\begin{eqnarray}}
\def\eea{\end{eqnarray}}
\def\ba{\begin{eqnarray}}
\def\ea{\end{eqnarray}}

\def\tree{{\rm tree}}

\newbox\charbox
\newbox\slabox
\def\s#1{{      
        \setbox\charbox=\hbox{$#1$}
        \setbox\slabox=\hbox{$/$}
        \dimen\charbox=\ht\slabox
        \advance\dimen\charbox by -\dp\slabox
        \advance\dimen\charbox by -\ht\charbox
        \advance\dimen\charbox by \dp\charbox
        \divide\dimen\charbox by 2
        \raise-\dimen\charbox\hbox to \wd\charbox{\hss/\hss}
        \llap{$#1$} }}

\begin{document}

\title{Three-Loop Superfiniteness of $\NeqEight$ Supergravity}

\author{Z.~Bern${}^a$, J.~J.~Carrasco${}^a$, L.~J.~Dixon${}^{b}$,
 H.~Johansson${}^a$, 
D.~A.~Kosower${}^{c,d}$ and  R.~Roiban${}^e$ }
\affiliation{
${}^a$Department of Physics and Astronomy, UCLA, Los Angeles, CA
90095-1547, USA  \\
${}^b$Stanford Linear Accelerator Center,
              Stanford University,
             Stanford, CA 94309, USA \\
${}^c$Service de Physique Th\'eorique, CEA--Saclay,
          F--91191 Gif-sur-Yvette cedex, France\\
${}^d$Institut f\"ur Theoretische Physik, Universit\"at Z\"urich,
        CH--8057  Z\"urich, Switzerland         \\
${}^e$Department of Physics, Pennsylvania State University,
           University Park, PA 16802, USA
}

\begin{abstract}
We construct the three-loop four-point amplitude of $\NeqEight$
supergravity using the unitarity method.  The amplitude is ultraviolet
finite in four dimensions.  Novel cancellations, not predicted by
traditional superspace power-counting arguments, render its degree of
divergence in $D$ dimensions no worse than that of $\NeqFour$
super-Yang-Mills theory --- a finite theory in four dimensions.
Similar cancellations can be identified at all loop orders in certain
unitarity cuts, suggesting that $\NeqEight$ supergravity may be a
perturbatively finite theory of quantum gravity.
\end{abstract}

\pacs{04.65.+e, 11.15.Bt, 11.25.Db, 12.60.Jv \hspace{1cm}}

\maketitle


While physicists do not yet know how to construct an
ultraviolet-finite, point-like quantum field theory of gravity
in four dimensions, neither have they shown that such a construction 
is impossible.  Point-like theories of gravity are non-renormalizable,
because the gravitational coupling is dimensionful.  To date, no known
symmetry has proven capable of taming the divergences, leading to the
widespread belief that all such theories require new physics in the
ultraviolet (UV).  These beliefs were historically an important motivation
for the development of string theory. Were a finite 
four-dimensional point-like theory of gravity to be found,
surely either a new symmetry or non-trivial dynamical mechanism must
underpin it.  The discovery of either would have a 
fundamental impact on our understanding of gravity.

Supersymmetry has been studied extensively as a mechanism for taming
UV divergences (see {\it e.g.}
refs.~\cite{HLK,HoweStelleReview,HoweStelleNew}).  
Although assumptions about the existence
of different types of superspaces lead to different power counting,
any superspace argument delays the onset of divergences only by a
limited number of loops.  For example, pure minimal (${\cal N}=1$) 
supergravity cannot diverge until at least three 
loops~\cite{Grisaru,DeserKayStelle}.  
For maximal $\NeqEight$ supergravity~\cite{CremmerJuliaScherk},
were a fully covariant superspace to exist, divergences would be 
delayed until seven loops.  With the additional
(unconventional) assumption that all fields respect ten-dimensional
general coordinate invariance, one can even delay the divergence to
nine loops~\cite{KellyPrivate}.  Recent arguments~\cite{GreenII} 
using the type~II string non-renormalization theorems of 
Berkovits~\cite{Berkovits} suggest that divergences in the corresponding 
supergravity theory may indeed not arise before this loop order, 
though issues with smoothness in the low-energy limit do weaken 
this prediction~\cite{GreenII}.  Beyond
this order, no known purely supersymmetric mechanism can avoid divergences.
String dualities also hint at UV finiteness for 
$\NeqEight$ supergravity~\cite{DualityArguments}, 
unless the situation is spoiled by towers of light nonperturbative 
states from branes wrapped on the compact dimensions~\cite{OoguriPrivate}.

Nonetheless, a different line of reasoning~\cite{BDDPR} using the
unitarity method~\cite{UnitarityMethod} has provided direct evidence
that $\NeqEight$ supergravity may be UV finite to {\it all\/}
loop orders~\cite{Finite}.  (See also ref.~\cite{KITPTalk}.)  At one
loop, all known multi-graviton amplitudes in the theory 
(including all with up to six gravitons) can be expressed
solely in terms of scalar box integrals; neither triangle nor
bubble integrals appear~\cite{OneloopMHVGravity,NoTriangle}.
Supersymmetry, factorization and infrared arguments provide strong
evidence that the same is true for all one-loop amplitudes.
This ``no-triangle hypothesis''~\cite{NoTriangle} 
implies a set of surprising
cancellations which go beyond any known superspace argumentation.  
Generalized unitarity cuts, isolating one-loop
subamplitudes inside higher-loop amplitudes, then imply 
specific multi-loop cancellations~\cite{Finite}.  
Are similar cancellations present
in all contributions to multi-loop amplitudes, and do they 
render the theory UV finite?

In this paper, we take a concrete step toward addressing these questions by
presenting the complete three-loop four-point amplitude of $\NeqEight$
supergravity.  Details of the computation will appear
elsewhere~\cite{Forthcoming}.  Here we show that the amplitude 
possesses the cancellations expected if the theory were indeed finite to 
all loop orders.

Ref.~\cite{BDDPR} analyzed iterated two-particle cuts
to all loop orders, and argued that $\NeqEight$ supergravity is finite
for
\begin{equation}
D< {10 / L} + 2  \hskip 1 cm  (L>1)\,,
\label{OldPowerCount}
\end{equation}
where $L$ is the loop order and $D$ is the dimension.  (For $L=1$, the
finiteness bound is $D<8$, not $D<12$.)

A similar analysis for $\NeqFour$ super-Yang-Mills
theory~\cite{BRY,BDDPR}, gives the finiteness condition,
\begin{equation}
D < {6/ L} + 4  \hskip 1 cm (L > 1) \,.
\label{SuperYangMillsPowerCount}
\end{equation}
The bound~(\ref{SuperYangMillsPowerCount}) differs somewhat from earlier
superspace power counting~\cite{HoweStelleYangMills}, although all
bounds confirm UV finiteness of $\NeqFour$ super-Yang-Mills theory in
$D=4$. The bound (\ref{SuperYangMillsPowerCount}) has been proven to
all orders~\cite{HoweStelleNew} using ${\cal N}= \nobreak 3$ harmonic
superspace~\cite{HarmonicSuperspace}. Explicit computations show that 
it is saturated through four
loops~\cite{BRY,BDDPR,ThreeFourLoop,Finite}.

The $\NeqEight$ supergravity bound~(\ref{OldPowerCount}) corresponds,
in the language of effective actions, to a one-particle irreducible
effective action starting with loop integrals multiplied by ${\cal
D}^4 R^4$ at each loop order beyond $L=1$.  Here $R^4$ is a shorthand for the
supersymmetrization of a particular contraction of four Riemann
tensors~\cite{DeserKayStelle}, and ${\cal D}$ denotes a generic
covariant derivative. 
The stronger, ``superfinite'' bound (\ref{SuperYangMillsPowerCount}),
if applied to $\NeqEight$
supergravity, would differ from \eqn{OldPowerCount}
beginning at $L=3$ for general
$D$, although both bounds imply three-loop finiteness for $D=4$.
It corresponds to a three-loop effective action beginning with
${\cal D}^6 R^4$, not ${\cal D}^4 R^4$.  As the supergravity
finiteness bound (\ref{OldPowerCount}) is based on only a limited set
of unitarity cuts~\cite{BDDPR}, additional (stronger) cancellations
may be missed~\cite{Finite}.


\begin{figure}[t]
\centerline{\epsfxsize 3. truein \epsfbox{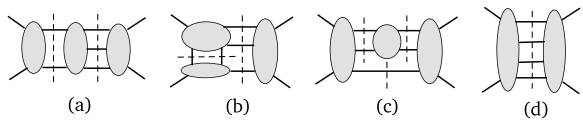}}
\caption[a]{\small Generalized cuts used to determine the 
three-loop four-point amplitude.}
\label{CutsFigure}
\end{figure}

To study this issue, we use the unitarity
method~\cite{UnitarityMethod,BRY} to build the three-loop four-point
$\NeqEight$ supergravity amplitude.  In this method, on-shell tree
amplitudes suffice as ingredients for computing amplitudes at any loop
order.  The reduction to tree amplitudes is crucial.  It allows the use of
the Kawai-Lewellen-Tye (KLT)~\cite{KLT} tree-level relations between
gravity and gauge theory amplitudes~\cite{BDDPR}, effectively 
reducing gravity computations to gauge theory ones.  
The original KLT relations express tree-level closed-string scattering 
amplitudes in terms of pairs of open-string ones.  
The perturbative massless states of the closed and open type~II superstring
compactified to four dimensions on a torus,
are those of $\NeqEight$ supergravity and $\NeqFour$
super-Yang-Mills theory, respectively.
Thus, in the limit of energies well below the string scale, 
the KLT relations express $\NeqEight$ supergravity tree amplitudes 
as quadratic combinations of $\NeqFour$ super-Yang-Mills tree
amplitudes (see {\it e.g.} ref.~\cite{OneloopMHVGravity}).
At tree level there are no subtleties in taking this limit.  

We use the generalized unitarity cuts~\cite{GeneralizedUnitarity}
illustrated in \fig{CutsFigure}. Together with the iterated
two-particle cuts evaluated in refs.~\cite{BRY,BDDPR}, these cuts
completely determine any massless three-loop four-point amplitude.
Since we are interested in the UV behavior of the amplitudes in $D$
dimensions, the unitarity cuts must be evaluated in $D$
dimensions~\cite{DDimUnitarity}. This renders the calculation more
difficult, because powerful four-dimensional spinor methods cannot be
used.  Some of the $D$-dimensional complexity is avoided
by performing internal-state sums in terms of the simpler on-shell gauge
supermultiplet of $D=10,\, {\cal N}=1$ super-Yang-Mills theory instead
of the $D=4,\, \NeqFour$ multiplet.  We have also performed various
four-dimensional cuts, which in practice provide a very useful guide.

\begin{figure}[t]
\centerline{\epsfxsize 3.0 truein \epsfbox{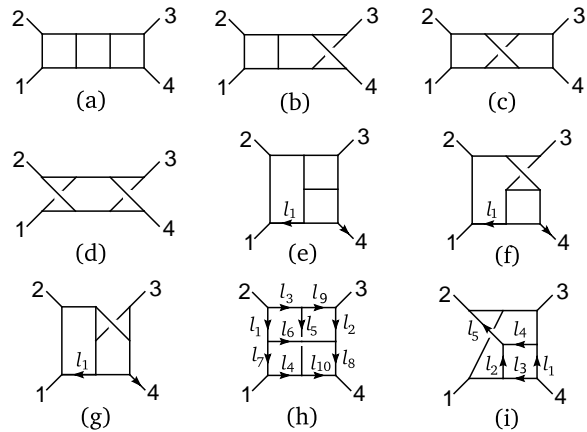}}
\caption[a]{\small Loop integrals appearing in both
$\NeqFour$ gauge-theory and $\NeqEight$ supergravity
three-loop four-point amplitudes.  The integrals are
specified by combining the diagrams' propagators with 
numerator factors given in \tab{NumeratorTable}.}
\label{DiagramsFigure}
\end{figure}

Our computation proceeds in two stages.  In the first stage we deduce the
three-loop $\NeqFour$ super-Yang-Mills amplitudes from generalized cuts,
including cuts (a)-(c) in \fig{CutsFigure}, and the iterated
two-particle cuts analyzed in refs.~\cite{BRY,BDDPR}. 
From the cuts we obtain a loop-integral representation of the
amplitude.  The diagrams in \fig{DiagramsFigure} describe 
the scalar propagators for the loop integrals. 
The numerator factor for each integral in the super-Yang-Mills case 
is given in the second column of \tab{NumeratorTable}.

In the second stage we use the KLT relations to write the cuts of the
$\NeqEight$ supergravity amplitude as sums over products of pairs of
cuts of the corresponding $\NeqFour$ super-Yang-Mills amplitude, 
including twisted non-planar contributions.
The iterated two-particle cuts studied in ref.~\cite{BDDPR},
together with the cuts in \fig{CutsFigure} evaluated here, suffice 
to fully reconstruct the supergravity amplitude.  
We find that the three-loop four-point $\NeqEight$ supergravity 
amplitude in $D$ dimensions is,
\begin{eqnarray}
M_4^{(3)} \! & = & \!\Bigl({\kappa \over 2}\Bigr)^8
  \! s t u M_4^\tree  \sum_{S_3}\,
\Bigl[ I^{\rm (a)} +  I^{\rm (b)} + {\textstyle {1\over 2}} I^{\rm (c)} 
  +  {\textstyle {1\over 4}} I^{\rm (d)}\nn \\
&& \null 
 + 2 I^{\rm (e)} 
 + 2 I^{\rm (f)} + 4 I^{\rm (g)} + 
   {\textstyle {1\over 2}}  I^{\rm (h)} 
 + 2 I^{\rm (i)} 
\Bigr] \,,  \hskip .3 cm 
\label{ThreeLoopAmplitude}
\end{eqnarray}
where $S_3$ represents the six independent permutations of legs
$\{1,2,3\}$, $\kappa$ is the gravitational coupling, and $M_4^\tree$ is the
supergravity four-point tree amplitude. The $I^{(x)}(s,t)$ are
$D$-dimensional loop integrals
corresponding to the nine diagrams in \fig{DiagramsFigure}, 
with numerator factors given in the third column of
\tab{NumeratorTable}. The Mandelstam invariants are $s = (k_1 +
k_2)^2$, $t=(k_2+k_3)^2$, $u=(k_1+k_3)^2$.  The numerical coefficients
in front of each integral in \eqn{ThreeLoopAmplitude} are symmetry 
factors of the diagrams.  Remarkably, 
the number of dimensions appears explicitly only
in the loop integration measure.

We remark that the amplitude~(\ref{ThreeLoopAmplitude})
could be used to study $D=11,\, {\cal N}=1$ supergravity compactified
on a circle or two-torus at three loops, just as the two-loop 
amplitude~\cite{BDDPR} was analyzed in ref.~\cite{GreenVanhove}.
That analysis, along with the assumption
that $M$-theory dualities hold at this loop order,
restricts the type~II string effective action
at three loops to start with ${\cal D}^6 R^4$, not ${\cal D}^4 R^4$.

\begin{table*}
\caption{The numerator factors of the integrals $I^{(x)}$ in
\fig{DiagramsFigure}. The first column labels the integral, the second
column the relative numerator factor for $\NeqFour$ super-Yang-Mills,
the third column the factor for $\NeqEight$ supergravity.  In the
Yang-Mills case an overall factor of $s t A_4^\tree$ has been removed,
while in the supergravity case an overall factor of $s t u M_4^\tree$
has been removed.  The loop momenta $l_i$ are the momenta of the
labeled propagators in \fig{DiagramsFigure}, and 
$l_{i,j}^2 = (l_i + l_j)^2$.
\label{NumeratorTable} }
\vskip .4 cm
\begin{tabular}{||c|c|c||}
\hline
Integral $I^{(x)}$ & $\NeqFour$ Super-Yang-Mills & $\NeqEight$ Supergravity  \\
\hline
\hline
(a)--(d) &  $s^2$ & $\vphantom{\Bigr|} [s^2]^2$   \\
\hline
(e)--(g) &  $s (l_1 + k_4)^2$ & 
        $\vphantom{\Bigr|} [s \, (l_1 + k_4)^2]^2$   \\
\hline
(h)&   $\; s l_{1,2}^2 + t l_{3,4}^2 - s l_5^2 - t l_6^2 - s t \; $ 
       & $\;(s l_{1,2}^2 + t l_{3,4}^2 - s t)^2
          - s^2 (2 (l_{1,2}^2 - t) + l_5^2 ) l_5^2 
          - t^2 (2 (l_{3,4}^2 - s) + l_6^2)  l_6^2 
         \; $  \\
 & $$      
       & $ \; \null
      - s^2 (2 l_1^2 l_8^2 + 2 l_2^2 l_7^2 + l_1^2 l_7^2 + l_2^2 l_8^2)
       \vphantom{\bigl|_{A_A}} - t^2 (2 l_3^2 l_{10}^2 + 2 l_4^2 l_9^2 +
            l_3^2 l_9^2 + l_4^2 l_{10}^2 )
           + 2 s t l_5^2 l_6^2\;
         $
           \\
\hline
(i) & $\;s l_{1,2}^2 - t l_{3,4}^2 - {1\over 3} (s - t) l_5^2 \;$
   &  
 $( s l_{1,2}^2 - t l_{3,4}^2 )^2 
  \vphantom{\bigl|_{A_A}} \null - ( s^2 
     l_{1,2}^2 + t^2 l_{3,4}^2 
             + {1\over3} s t u ) l_5^2 $ \\
\hline
\end{tabular}
\end{table*}

With the explicit expression for the amplitude
(\ref{ThreeLoopAmplitude}) in hand, we may determine
the UV behavior straightforwardly.  In the super-Yang-Mills
case, the entries in the second column in \tab{NumeratorTable} contain
no more than two powers of loop momenta.
Accounting for the ten propagators of each diagram in \fig{DiagramsFigure}
and the three-loop integration measure,
we see that each integral separately satisfies the known 
super-Yang-Mills finiteness bound~(\ref{SuperYangMillsPowerCount}).

In contrast, the supergravity numerators, as given in the third column of
\tab{NumeratorTable}, contain up to four powers of loop momenta.
Separately, these integrals satisfy the bound
(\ref{OldPowerCount}). The iterated two-particle cuts evaluated in
ref.~\cite{BDDPR} give the integrals (a)-(g) in \fig{DiagramsFigure} and
\tab{NumeratorTable}.  All such contributions have numerator factors
which are squares of the Yang-Mills ones.  The entries (h) and (i) are
new and do not have this structure.

\begin{figure}[t]
\centerline{\epsfxsize 3. truein \epsfbox{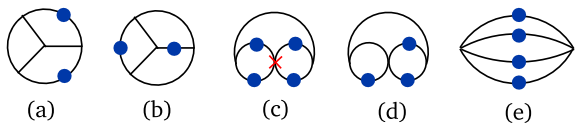}}
\caption[a]{\small The vacuum diagrams $V^{(x)}$ encoding the leading
UV behavior of the individual $\NeqEight$ supergravity
diagrams.  Dots on propagators represent squared propagators.  The
shaded cross in diagram (c) represents a numerator factor of $l^2$,
where $l$ is the momentum of a collapsed vertical propagator.}
\label{VacuumFigure}
\end{figure}

Might there be additional cancellations between the integrals in the 
$\NeqEight$ supergravity case? To check
this, we expand each integral in a series in the
external momenta, keeping only the leading terms. We thereby
keep terms with maximal powers of loop momenta and
set the external momenta to
zero in the propagators. Each integral reduces to a sum of vacuum
diagrams, possibly with squared propagators. 
\Fig{VacuumFigure} shows the resulting vacuum diagrams $V^{(x)}$.
Integrals (a)-(d) in \fig{DiagramsFigure} have no
powers of loop momenta in their numerators, and hence do not contribute to
the leading UV behavior.  The remaining integrals in 
\eqn{ThreeLoopAmplitude} reduce as follows,
\begin{eqnarray}
2 I^{\rm (e)} &\rightarrow&  4 V^{\rm(a)}\,, \hskip .5 cm 
2 I^{\rm(f)} \rightarrow  4 V^{\rm (b)}\,, \hskip .5 cm 
4 I^{\rm(g)} \rightarrow  8 V^{\rm(a)}\,, \nn\\
{\textstyle {1\over 2}}
I^{\rm(h)} &\rightarrow&  - 4 V^{\rm(a)} - 8 V^{\rm(b)} - 4 V^{\rm(c)} 
                  -  2 V^{\rm(e)}\,, \nn\\
2 I^{\rm(i)} &\rightarrow& - 8 V^{\rm(a)} + 4 V^{\rm(b)} + 8 V^{\rm(d)}\,,
\end{eqnarray}
taking into account the permutation sum over external legs and
suppressing an overall factor of $(s^2 + t^2 + u^2) s t u M_4^\tree$.
Using a momentum-conservation identity,
%
\begin{equation}
V^{\rm (c)} = 2 V^{\rm (d)}  - {\textstyle {1\over 2}} V^{\rm (e)},
\end{equation}
%
to eliminate $V^{\rm (c)}$,
{\it the coefficients of the remaining four vacuum diagrams 
cancel completely}.

Lorentz covariance implies that contributions with three powers
of loop momenta in the numerator are no more divergent
than integrals with only two powers.  We
have also found a rearrangement of the loop-momentum integrands
which makes manifest this quadratic behavior, equivalent
to the amplitude behaving as ${\cal D}^6 R^4$~\cite{Forthcoming}.  
Thus the $\NeqEight$
supergravity amplitude satisfies the same finiteness bound
(\ref{SuperYangMillsPowerCount}) at $L=3$ as the corresponding
$\NeqFour$ super-Yang-Mills amplitude.

\begin{figure}[t]
\centerline{\epsfxsize 1.4 truein \epsfbox{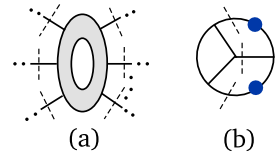}}
\caption[a]{\small A generalized cut (a) isolating a one-loop
subamplitude in an $L$-loop amplitude.  If a leg is external to the
entire amplitude, it should not be cut.  From the generalized cut (b)
we see that diagram (a) in \fig{VacuumFigure} must cancel, since it
has a one-loop triangle subdiagram.}
\label{OneLoopCutFigure}
\end{figure}

Some of these cancellations have an all-loop generalization that 
can be understood as a direct consequence of the no-triangle
hypothesis~\cite{Finite}. Using generalized unitarity we may isolate
all one-loop subamplitudes of an $L$-loop amplitude as shown in
\fig{OneLoopCutFigure}(a).  For example, the cut shown in
\fig{OneLoopCutFigure}(b) rules out the appearance of the vacuum
diagram $V^{\rm (a)}$, because it would imply the appearance of a
triangle integral at one loop. Similarly, we can infer a cancellation
between vacuum diagrams $V^{\rm (c)}$ and $V^{\rm (d)}$.  (A squared
propagator counts as two sides of a triangle integral.)  At higher
loops, the generalized cut in \fig{OneLoopCutFigure}(a), together with
the no-triangle hypothesis, implies that any leading-singularity
vacuum diagram containing a triangle subdiagram must have a vanishing
coefficient.  However, this argument does not suffice to rule out
vacuum diagrams $V^{\rm (b)}$ and $V^{\rm (e)}$, because they have no
triangle subdiagrams.  Their coefficients nonetheless vanish,
demonstrating the existence of cancellations {\it beyond} those
implied by the no-triangle hypothesis.


This paper establishes through three loops that the four-point
amplitudes of $\NeqEight$ supergravity have the same ultraviolet
critical dimension~(\ref{SuperYangMillsPowerCount}) as the
corresponding $\NeqFour$ gauge-theory ones.  Fourteen powers of loop
momentum are extracted from the numerators of the three-loop
integrals.  This result is consistent with the manifest symmetries of
an off-shell ${\cal N} =7$ harmonic superspace~\cite{HoweStelleNew},
whose existence would imply UV finiteness for $D<12/L +2$.  However,
the cancellations we find go beyond this: generalized unitarity will
propagate the additional three-loop cancellations, as well as the
one-loop no-triangle constraint, into novel cancellations eliminating
increasing powers of loop momenta at {\it all\/} loop
orders~\cite{Forthcoming}.

To unravel the origin of these cancellations, and to constrain
potential superspace explanations, it is important to
compute additional $\NeqEight$ amplitudes.  Using the
unitarity method it should be feasible to compute the four- and
five-loop four-point amplitudes, as well as the two-loop five-point
amplitude. It should also be
possible to carry out refined all-order studies,
given the recursive nature of the formalism.  In particular, 
it is important to investigate the classes of contributions not
directly constrained by generalized unitarity and the no-triangle hypothesis.

The result presented here, in conjunction with the
all-loop-order evidence from unitarity~\cite{Finite} and string
theory hints of additional cancellations~\cite{DualityArguments,
Berkovits, GreenII},  points strongly towards the ultraviolet finiteness of 
$\NeqEight$ supergravity

We thank N.~Berkovits, E.~D'Hoker, M.~Green, C.~Hull, H.~Nicolai,
H.~Ooguri, J.~Schwarz, K.~Stelle and P.~Vanhove for valuable
discussions.  We especially thank M.~Perelstein for initial
collaboration many years ago.  This research was supported by the US
Department of Energy under contracts DE--FG03--91ER40662 and
DE--AC02--76SF00515, and the National Science Foundation under grant
PHY-0608114.


\end{document}